\documentclass[11pt,a4paper]{article}

\usepackage[margin=1in]{geometry}
\usepackage{amsmath,amssymb,amsfonts,bm}
\usepackage{graphicx}
\usepackage{enumitem}
\usepackage{xcolor}
\usepackage{tikz}

\usepackage[
    colorlinks=true,
    linkcolor=blue!60!black,
    citecolor=blue!60!black,
    urlcolor=blue!60!black
]{hyperref}

\usetikzlibrary{arrows.meta,decorations.pathreplacing,positioning,calc}
\definecolor{xcol}{RGB}{220,50,47}
\definecolor{zcol}{RGB}{38,139,210}
\definecolor{ycol}{RGB}{133,153,0}

\newcommand{\ket}[1]{\left|#1\right\rangle}
\newcommand{\bra}[1]{\left\langle#1\right|}

\newcommand{\sx}{\sigma_x}
\newcommand{\sy}{\sigma_y}
\newcommand{\sz}{\sigma_z}

\newcommand{\Tr}{\operatorname{Tr}}
\newcommand{\diag}{\operatorname{diag}}
\newcommand{\SO}{\mathrm{SO}}
\newcommand{\OO}{\mathrm{O}}

\newcommand{\conc}{\mathcal{C}}

\newcommand{\ahat}{\hat{\bm a}}
\newcommand{\bhat}{\hat{\bm b}}
\newcommand{\ez}{\bm e_z}

\graphicspath{{figures/}{./}}

\title{\textbf{Roto-Reflection Geometry of Pure Two-Qubit Entanglement}}
\author{Stanislav Filatov and Marcis Auzinsh \\ University of Latvia}
\date{October 22, 2026}

\begin{document}

\maketitle

\begin{abstract}
Pure two-qubit entanglement is usually characterized by scalar quantities such as concurrence. Here we show that it also has a natural geometric form. In the Pauli correlation tensor, maximally entangled states appear as improper orthogonal maps between two local Bloch spheres. These maps are roto-reflections. For partially entangled pure states, the same roto-reflection geometry is recovered after separating the contraction associated with concurrence. We call the corresponding geometric object the Entanglement Roto-Reflection Plane (ERRP). It organizes the maximally correlated directions of the two-qubit state and provides a covariant geometric complement to the scalar magnitude of entanglement. We further illustrate the usefulness of this geometry through partial transposition, where it provides a direct construction of the maximally entangled negative eigenvector and a geometric interpretation of negativity.
\end{abstract}

\section{Introduction}

Imagine that a quantum linguist, perhaps with suspiciously Wittgensteinian habits, enters the lab with a simple question: what do the words ``same'' and ``different'' mean for two qubits?

For two ordinary arrows on a sphere, the answer seems harmless. If the arrows point in the same direction, call them the same. If they point in opposite directions, call them different. Measure both qubits along that direction, and the outcomes obediently agree or disagree. The dictionary works.

But only along that direction. Two qubits that are perfectly different along \(z\) may look only half-different along \(x\). The word has slipped. It was not attached to the pair of qubits after all, but to the axis we chose to ask about.

The singlet state fixes this. Measure the two qubits along any common axis, and the outcomes are always opposite. The singlet does not merely say ``different along \(z\)''. It says ``different'', full stop.

Now comes the tempting mirror question: can two qubits say ``same'' with the same force? Can they agree along every axis in the same way that the singlet disagrees along every axis? Surprisingly, no. The geometry of entanglement refuses this perfectly symmetric dictionary. Somewhere, something has to be reflected. This work is about revealing that mirror and making its geometry explicit.

In a two-qubit state, the essential information is not always carried by the local Bloch vectors. For maximally entangled states the reduced states are maximally mixed, so the individual Bloch vectors vanish, while the correlations remain highly structured. To see entanglement geometrically, one must therefore find a way to draw not only local states, but the organization of correlations between possible measurement directions.

Many geometric languages for two-qubit states have been developed with this aim, including Hopf-fibration descriptions of pure two-qubit states, Geometric Algebra models, generalized Bloch-vector representations, and toroidal geometries \cite{mosseri2001,havel2004,wie2020,rau2021symmetries,jakobczyk2001geometry,chang2024geometric}. Other approaches retain the local Bloch-sphere language while adding structures that encode correlations, such as the Pauli or Fano decomposition, local-unitary invariants of the correlation matrix, Lorentz-normal forms under local filtering, steering ellipsoids, and Bloch-matrix visualizations \cite{makhlin2002,verstraete2001,devicente2007,jevtic2014,milne2014,gamel2016}.

There are two works that are the closest in spirit to the present manuscript. The first is our previous two-Bloch-sphere representation of pure two-qubit states and unitaries \cite{filatov2024twobloch}. There, the visualization used a nested two-sphere structure: local Bloch vectors occupied the inner spheres, while entanglement was represented through the relative organization of coordinate frames in the outer shells. Maximally entangled states appeared through a reversal of handedness between the two frames, and Geometric Algebra supplied the language of planes of rotation needed to describe the corresponding transformations.

The second is Gamel's entangled-Bloch-sphere representation \cite{gamel2016}. Gamel uses the Bloch matrix and the singular value decomposition of its correlation block to draw two Bloch spheres with paired principal correlation axes. He also uses this decomposition to study positivity, local and nonlocal unitary transformations, invariants, entanglement criteria, and regions of two-qubit state space. For the present work, the key point is that the correlation tensor emerges as a natural source of two-sphere geometry.

The present paper combines this tensor-based starting point with the nested visual organization of our earlier work. Local Bloch vectors remain the inner, ordinary Bloch-sphere objects, while entanglement is represented as an additional geometric layer. The question we ask is not only which principal axes are correlated, but what geometric map is defined by the full Pauli correlation tensor.

We call the resulting object the Entanglement Roto-Reflection Plane, or ERRP. It is not another scalar entanglement measure; pure-state entanglement already has such a measure in concurrence \cite{wootters1998}. Instead, the ERRP describes the form of the correlations. For maximally entangled states it makes the orientation-reversing geometry of the correlation tensor explicit as a real plane-and-angle structure. For partially entangled pure states, it shows how the same underlying geometry remains after the contraction associated with concurrence is separated. The goal of the paper is to make this structure precise, computable, and drawable.

\section{Pauli decomposition and two-Bloch-sphere visualization}

Let \(\sigma_0=I\) and let \(\sigma_1,\sigma_2,\sigma_3\) denote the Pauli matrices. Any two-qubit density matrix can be expanded as
\begin{equation}
    \rho
    =
    \frac{1}{4}
    \sum_{\mu,\nu=0}^{3}
    R_{\mu\nu}\,
    \sigma_\mu\otimes\sigma_\nu,
    \qquad
    R_{\mu\nu}
    =
    \Tr\left(\rho\,\sigma_\mu\otimes\sigma_\nu\right).
    \label{eq:pauli-decomposition}
\end{equation}
The real coefficients \(R_{\mu\nu}\) may be arranged as a \(4\times4\) Pauli coefficient matrix,
\begin{equation}
    R
    =
    \begin{pmatrix}
        1 & \bm b^T\\
        \bm a & T
    \end{pmatrix}.
    \label{eq:pauli-matrix-block}
\end{equation}
Here \(\bm a\) and \(\bm b\) are the reduced Bloch vectors of the two qubits, while
\begin{equation}
    T_{ij}
    =
    \Tr\left(\rho\,\sigma_i\otimes\sigma_j\right),
    \qquad i,j\in\{x,y,z\},
    \label{eq:correlation-tensor-trace}
\end{equation}
is the \(3\times3\) correlation tensor. For pure states, which are the focus of this work, this reduces to
\begin{equation}
    T_{ij}
    =
    \bra{\psi}
    \sigma_i\otimes\sigma_j
    \ket{\psi}.
    \label{eq:correlation-tensor-pure}
\end{equation}
Thus the same Pauli decomposition gives the local vectors \(\bm a,\bm b\) and the correlation tensor \(T\) in one step.

For local measurement directions \(\bm x,\bm y\in\mathbb{R}^3\), the joint correlation is
\begin{equation}
    E(\bm x,\bm y)
    =
    \bra{\psi}
    (\bm x\cdot\bm\sigma)\otimes(\bm y\cdot\bm\sigma)
    \ket{\psi}
    =
    \bm x^T T\bm y.
    \label{eq:joint-correlation}
\end{equation}
This formula is the starting point for the geometry. Given a direction \(\bm y\) on Bob's Bloch sphere, the vector \(T\bm y\) is the correlation vector on Alice's side. When \(T\bm y\neq0\), the Alice direction that maximizes the correlation with Bob's direction \(\bm y\) is parallel to \(T\bm y\). The correlation tensor can therefore be read as a direction-pairing rule between the two Bloch spheres.

This block structure gives the visualization its natural layers. The first row and first column contain local Bloch-vector information. The lower-right \(3\times3\) block contains two-qubit correlations. In the visualization used here, the reduced local state is represented by an inner Bloch vector, while correlation directions are displayed in an outer shell.

\begin{figure}
    \centering
\begin{tikzpicture}
\node[
    rounded corners=8pt,
    inner sep=0pt,
    clip
] {
    \includegraphics[width=0.96\textwidth]{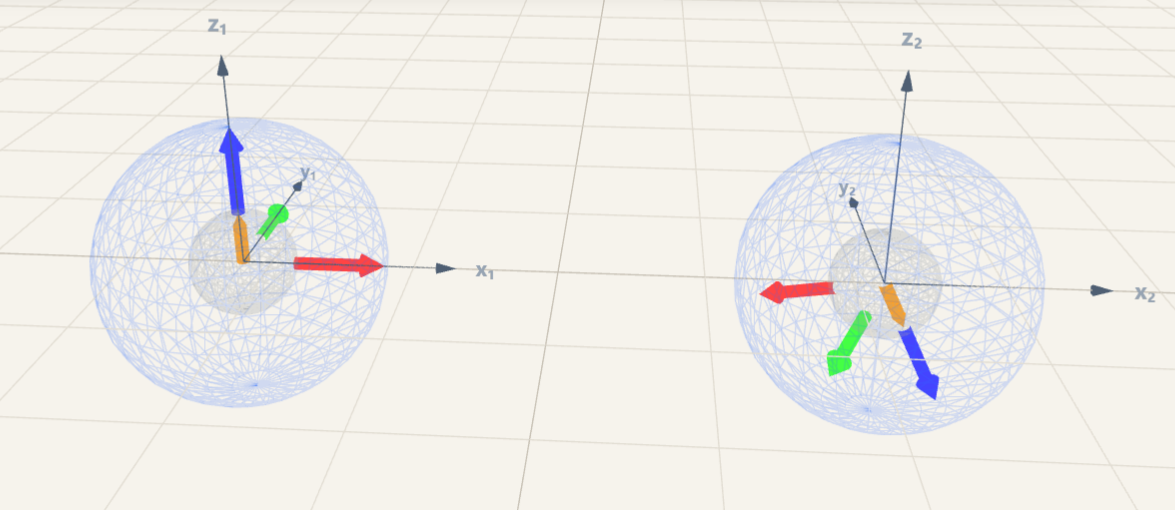}
};
\end{tikzpicture}
\caption{Arbitrary pure partially entangled two-qubit state. Inner spheres with reduced Bloch vectors represent the local reduced-state information while outer shells with correlated triads represent entangled part of the state. Arrows of the same color represent maximally correlated directions.}
\label{Fig1}
\end{figure}

There are two useful ways to display this outer shell. In the triad view (Fig. \ref{Fig1}), three paired directions are shown simultaneously, giving a local frame-like picture of the correlation structure. In the single-arrow view, one direction is chosen on one Bloch sphere and the corresponding maximally correlated direction is shown on the other. This is a small but useful refinement: it makes the tensor readable as a continuous direction-pairing rule rather than only as a relation between selected axes. The use of a common coordinate frame on both spheres also makes the correlation geometry directly comparable across the two subsystems.

\begin{figure}
\centering
\begin{tikzpicture}[scale=4.2,>=Stealth]

    \def\r{0.4}
    \pgfmathsetmacro{\cc}{sqrt(1-\r*\r)}

    \draw[->] (0,0) -- (1.08,0) node[below right] {$r$};
    \draw[->] (0,0) -- (0,1.08) node[above left] {$\conc$};

    \draw[thick] (1,0) arc[start angle=0,end angle=90,radius=1];
    \node at (0.7,0.95) {$r^2+\conc^2=1$};

    \draw[very thick,->] (0,0) -- (\r,\cc);
    \draw[dashed] (\r,\cc) -- (\r,0);
    \draw[dashed] (\r,\cc) -- (0,\cc);

\pgfmathsetmacro{\ang}{atan2(\cc,\r)}
\draw[thin] (0.18,0) arc[start angle=0,end angle=\ang,radius=0.18];
\node at ({0.29*cos(\ang/2)},{0.29*sin(\ang/2)}) {$2\theta$};

    \fill (\r,\cc) circle (0.45pt);

    \draw (0,1) -- +(-0.018,0) node[left] {$1$};

    \draw[decorate,decoration={brace,mirror,amplitude=4pt}]
        (0,0) -- (\r,0)
        node[midway,below=6pt] {$r$};

    \draw[decorate,decoration={brace,mirror,amplitude=4pt}]
        (\r,0) -- (1,0)
        node[midway,below=6pt] {$1-r$};

    \draw[decorate,decoration={brace,amplitude=4pt}]
        (0,0) -- (0,\cc)
        node[midway,left=6pt] {$\conc$};

    \node[below] at (0.14,-0.14) {\scriptsize local Bloch vector};
    \node[below] at ({0.05+(\r+1)/2},-0.14) {\scriptsize outer shell};

\end{tikzpicture}
\caption{Relation between the reduced Bloch-vector length $r$ and the concurrence $C$. The horizontal axis also indicates the remaining radial interval $1-r$ used for the outer shell in Fig.~1. The angle $2\theta$ corresponds to twice the Schmidt angle introduced in Sec.~4.}
\label{fig:r-c-projection}
\end{figure}

The shell structure shown in Fig.~1 reflects the complementarity between the local Bloch-vector length $r$ and the entanglement of a pure two-qubit state, quantified by the concurrence $C$. These quantities satisfy
\begin{equation}
    r^2+C^2=1.
\end{equation}

In our representation (Fig.~\ref{Fig1}), the inner Bloch vectors describe the reduced local states, while the outer shells represent entanglement, with radial width $1-r$. The relation between this width and the concurrence $C$ is illustrated in Fig.~\ref{fig:r-c-projection}, where $r$ and $C$ are perpendicular projections of a unit vector, satisfying $r^2+C^2=1$. Although the shell width $1-r$ is not equal to the concurrence, the connection becomes direct when using squared radial coordinates: the outer shell then corresponds to $1-r^2=C^2$. A variation of the visualization, in which the entanglement-related geometry is scaled directly by concurrence and placed between the two Bloch spheres, was recently proposed by Grosse-Holz~\cite{grosseholz2026}.

The triads shown in Fig.~\ref{Fig1} already contain the full directional information of the entangled shell. The next section extracts the underlying roto-reflection geometry that organizes them.

\section{Maximally entangled states: from reflection to ERRP}

We start with the maximally entangled state
\begin{equation}
    \ket{\Phi^+}
    =
    \frac{1}{\sqrt{2}}(\ket{00}+\ket{11}).
\end{equation}
Its Pauli correlations are
\begin{equation}
    \langle \sx\otimes\sx\rangle=1,
    \qquad
    \langle \sy\otimes\sy\rangle=-1,
    \qquad
    \langle \sz\otimes\sz\rangle=1,
\end{equation}
and all off-diagonal correlations vanish. Hence
\begin{equation}
    T_{\Phi^+}
    =
    \diag(1,-1,1).
    \label{eq:Tphi}
\end{equation}
We will also denote this canonical matrix by
\begin{equation}
    O_0 := T_{\Phi^+}.
    \label{eq:O0def}
\end{equation}
This matrix is not merely a list of correlation signs. It is a geometric transformation: reflection in the \(zx\)-plane. Equivalently,
\begin{equation}
    O_0^TO_0=I,
    \qquad
    \det O_0=-1.
\end{equation}
Thus the simplest Bell-state correlation tensor is already an improper orthogonal map.

In \(\ket{\Phi^+}\), directions in the \(zx\)-plane are correlated, while the direction perpendicular to the plane is anti-correlated. The plane itself is therefore the natural reflection plane of the entanglement correlations.

Every maximally entangled two-qubit state is locally equivalent to \(\ket{\Phi^+}\). Its correlation tensor therefore has the form
\begin{equation}
    T
    =
    R_A O_0 R_B^T,
    \label{eq:MEStransform}
\end{equation}
where \(R_A,R_B\in\SO(3)\) are the local Bloch-sphere rotations induced by local unitaries. Since \(R_A\) and \(R_B\) are proper rotations,
\begin{equation}
    T^TT=I,
    \qquad
    \det T
    =
    \det(R_A)\det(O_0)\det(R_B^T)
    =
    -1.
\end{equation}
Thus every maximally entangled state defines an improper orthogonal transformation between the two Bloch spheres. In three dimensions, such a transformation is a roto-reflection: a reflection in a plane, followed by a rotation about the normal to that plane.

The ERRP makes this roto-reflection explicit. For a maximally entangled state with correlation tensor \(T\neq -I\), the ERRP is the plane perpendicular to the real eigendirection \(\bm n\) satisfying
\begin{equation}
    T\bm n=-\bm n.
    \label{eq:eigen-normal-MES}
\end{equation}
The associated rotation angle \(\phi\) is determined by
\begin{equation}
    \cos\phi
    =
    \frac{\Tr(T)+1}{2}.
    \label{eq:phi-angle-MES}
\end{equation}
The vector \(\bm n\) determines the reflection plane. The sign of \(\bm n\) and the sign of the rotation angle are conventional: the pairs
\[
    (\bm n,\phi)
    \quad\text{and}\quad
    (-\bm n,-\phi)
\]
represent the same roto-reflection geometry. The complete ERRP geometry is therefore an oriented normal and signed angle modulo this equivalence.

There is one important degeneracy. For the singlet-type case \(T=-I\), every direction satisfies \(T\bm n=-\bm n\). The angle is \(\phi=\pi\), and the ERRP plane is not unique. Geometrically, this is point inversion: reflection in any plane followed by a \(\pi\)-rotation gives the same transformation. In the equivalent roto-inversion description, the same case is point inversion with zero additional rotation. Thus the singlet is a maximally symmetric limiting case rather than a failure of the construction.

\begin{figure}
    \centering
\begin{tikzpicture}
\node[
    rounded corners=8pt,
    inner sep=0pt,
    clip
] {
    \includegraphics[width=0.96\textwidth]{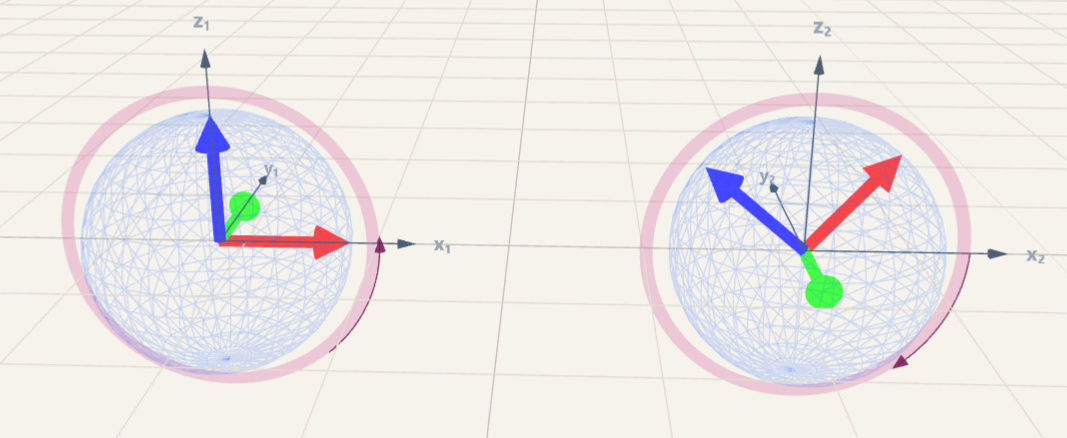}
};
\end{tikzpicture}
\caption{Maximally entangled state with ERRP equal to the \(zx\)-plane and a nonzero in-plane rotation. The state is obtained from \(\ket{\Phi^+}\) by a local rotation \(I\otimes e^{-i\alpha\sigma_y/2}\) of the second qubit. The arc indicates the rotation angle of the roto-reflection.}
\label{Fig2}
\end{figure}

In Fig.~\ref{Fig2} we show a maximally entangled state represented by paired triads together with its ERRP. Starting from \(\ket{\Phi^+}\), whose ERRP is the \(zx\)-plane, we apply a local rotation about the \(y\)-axis on the second qubit. This turns the pure reflection into a nontrivial roto-reflection while keeping the reflection plane easy to see. The direction of the angle arrow indicates the in-plane rotation needed, together with the reflection, to relate the two correlation triads.

\section{Partially entangled states: extracting the geometry}

We now turn to partially entangled pure states. Start with the Schmidt-form state
\begin{equation}
    \ket{\psi_\theta}
    =
    \cos\theta\,\ket{00}
    +
    \sin\theta\,\ket{11},
    \qquad
    0\leq \theta\leq \frac{\pi}{4}.
    \label{eq:schmidt}
\end{equation}
Its concurrence is
\begin{equation}
    \conc=\sin 2\theta.
    \label{eq:concurrence}
\end{equation}
The local Bloch vectors point along \(z\) and have length \(\cos 2\theta\). The correlation tensor is
\begin{equation}
    T_\theta
    =
    \diag(\conc,-\conc,1).
    \label{eq:Ttheta}
\end{equation}
This tensor is no longer orthogonal unless \(\conc=1\). Still, the reflection structure of \(\ket{\Phi^+}\) remains visible:
\begin{equation}
    T_\theta
    =
    \underbrace{\diag(1,-1,1)}_{O_0}
    \underbrace{\diag(\conc,\conc,1)}_{P_0}.
    \label{eq:schmidt-factorization}
\end{equation}
The first factor $O_0$ is the same improper orthogonal reflection as before. The second factor $P_0$ is a positive contraction: it leaves the reduced Bloch-vector direction unchanged and contracts the transverse plane by the concurrence.

The same statement can be written as
\begin{equation}
    P_0
    =
    \conc I+(1-\conc)\ez\ez^T.
    \label{eq:P0}
\end{equation}
The special direction \(\ez\) is the direction of the reduced Bloch vectors. Partial entanglement therefore does not remove the roto-reflection geometry; it contracts it. After the contraction is separated, the same canonical reflection \(O_0\) is recovered.

For a general pure state, the Schmidt form is rotated locally. Hence
\begin{equation}
    T
    =
    R_A T_\theta R_B^T,
    \label{eq:generalT}
\end{equation}
where \(R_A,R_B\in\SO(3)\). Using Eq.~\eqref{eq:schmidt-factorization},
\begin{equation}
    T
    =
    R_A O_0 P_0 R_B^T
    =
    (R_A O_0 R_B^T)(R_B P_0 R_B^T).
\end{equation}
This gives
\begin{equation}
    T=O P_B,
    \label{eq:right-polar}
\end{equation}
where
\begin{equation}
    O=R_AO_0R_B^T,
    \qquad
    P_B=R_B P_0 R_B^T.
    \label{eq:OPBdefs}
\end{equation}
The orthogonal factor satisfies
\begin{equation}
    O\in\OO(3),
    \qquad
    \det O=-1.
    \label{eq:Oimproper}
\end{equation}
Thus the orientation-reversing character of the entanglement geometry survives arbitrary local rotations.

Let
\[
    \bhat=R_B\ez,
    \qquad
    \ahat=R_A\ez.
\]
These are the directions of the reduced Bloch vectors for Bob and Alice, respectively, whenever \(0<\conc<1\). From Eq.~\eqref{eq:P0},
\begin{align}
    P_B
    &=
    R_B\left(\conc I+(1-\conc)\ez\ez^T\right)R_B^T\\
    &=
    \conc I+(1-\conc)\bhat\bhat^T.
    \label{eq:PB}
\end{align}
Similarly,
\begin{equation}
    P_A
    =
    \conc I+(1-\conc)\ahat\ahat^T.
    \label{eq:PA}
\end{equation}
The same tensor can also be written as
\begin{equation}
    T=P_AO.
    \label{eq:left-polar}
\end{equation}
This follows because \(O_0\) and \(P_0\) commute in Schmidt form. Indeed,
\[
    P_AO
    =
    R_A P_0 R_A^T R_A O_0 R_B^T
    =
    R_A P_0 O_0 R_B^T
    =
    R_A O_0 P_0 R_B^T
    =
    T.
\]

Combining these expressions gives the compact decomposition
\begin{equation}
    T
    =
    \conc O+(1-\conc)\ahat\bhat^T,
    \qquad
    0<\conc<1.
    \label{eq:main-decomposition}
\end{equation}
To see this, write
\[
    T
    =
    O\left(\conc I+(1-\conc)\bhat\bhat^T\right)
    =
    \conc O+(1-\conc)O\bhat\bhat^T,
\]
and note that
\[
    O\bhat
    =
    R_AO_0R_B^T R_B \ez
    =
    R_AO_0\ez
    =
    R_A\ez
    =
    \ahat.
\]

This is the pure-state factorization used in the rest of the paper. For maximally entangled states, \(\conc=1\) and the local Bloch vectors vanish; the factorization reduces to \(T=O\). For partially entangled pure states, \(0<\conc<1\), the tensor is not orthogonal, but it contains the same improper orthogonal factor \(O\). For product states, \(\conc=0\) and \(T\) has rank one; there is no entanglement roto-reflection geometry to extract.

The term \((1-\conc)\ahat\bhat^T\) in Eq.~\eqref{eq:main-decomposition} should not be confused with the product-correlation baseline \(\bm a\bm b^T\) determined by the two reduced states. The latter is obtained directly from the inner Bloch vectors on the two spheres. Equation~\eqref{eq:main-decomposition} instead separates the oriented correlation map into a concurrence-scaled improper geometry and the longitudinal completion required to reconstruct \(T\).

\begin{figure}
    \centering
\begin{tikzpicture}
\node[
    rounded corners=8pt,
    inner sep=0pt,
    clip
] {
    \includegraphics[width=0.96\textwidth]{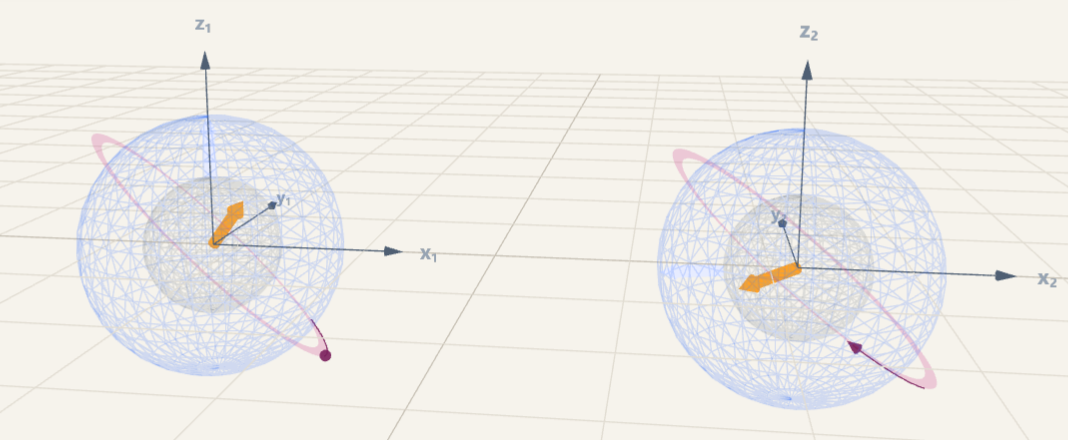}
};
\end{tikzpicture}
\caption{Representation of an arbitrary pure partially entangled state where entanglement geometry is described through an ERRP.}
\label{Fig3}
\end{figure}

In Fig.~\ref{Fig3} we return to an arbitrary pure partially entangled state, now represented using only the ERRP. It alone encodes the full correlation structure of the entangled shell, making the principal axis triads redundant. There is also a symmetry advantage. In the triad representation one sphere plays the role of reference and the other is the target, with directions mapped through $O$; as a consequence, local rotations on the reference sphere induce rotations of the triad on the target. The ERRP carries no such asymmetry - a local rotation on either sphere moves the plane, and the two spheres enter on equal footing.

What remains is a compact representation of any pure two-qubit state built from familiar objects: two Bloch vectors for the marginals, and a single new object, the ERRP, for the correlations.

\section{Extraction, magnitude, and covariance}

The full Pauli decomposition gives the quantities needed to extract the geometry. Starting from
\[
    \rho=\ket{\psi}\bra{\psi},
\]
compute
\begin{equation}
    R_{\mu\nu}
    =
    \Tr(\rho\,\sigma_\mu\otimes\sigma_\nu),
    \qquad
    \sigma_0=I,
\end{equation}
and arrange the result as
\begin{equation}
    R
    =
    \begin{pmatrix}
        1 & \bm b^T\\
        \bm a & T
    \end{pmatrix}.
    \label{eq:extraction-block}
\end{equation}
Thus \(\bm a\), \(\bm b\), and \(T\) are obtained from the same Pauli coefficient matrix.

For a maximally entangled state, \(T\) is already the ERRP geometry:
\begin{equation}
    O=T.
\end{equation}
For a partially entangled pure state, since the singular values of \(T\) for a pure state are
\[
    1,
    \quad
    \conc,
    \quad
    \conc,
\]
and since the orthogonal factor has determinant \(-1\),
\begin{equation}
    \det T=-\conc^2.
    \label{eq:detT}
\end{equation}
Therefore, for pure two-qubit states,
\begin{equation}
    \conc=\sqrt{-\det T}.
    \label{eq:Cfromdet}
\end{equation}

The determinant gives the scalar magnitude of entanglement; the improper orthogonal factor gives its roto-reflection geometry.

The local directions are
\begin{equation}
    \ahat=\frac{\bm a}{|\bm a|},
    \qquad
    \bhat=\frac{\bm b}{|\bm b|}.
\end{equation}
The improper orthogonal factor is then
\begin{equation}
    O
    =
    \frac{T-(1-\conc)\ahat\bhat^T}{\conc},
    \qquad
    0<\conc<1.
    \label{eq:Oextraction}
\end{equation}
Equivalently, \(O\) can be obtained from the ordinary Euclidean polar decomposition
\begin{equation}
    T=OP,
    \qquad
    P=(T^TT)^{1/2},
    \qquad
    O=TP^{-1},
    \label{eq:polar-decomposition}
\end{equation}
which is unique whenever \(T\) is invertible \cite{higham1986}. For pure entangled states this is the case precisely when \(\conc>0\).

Once \(O\) has been obtained, the ERRP is found by solving
\begin{equation}
    O\bm n=-\bm n.
    \label{eq:normaleq}
\end{equation}
The associated rotation angle is determined by
\begin{equation}
    \cos\phi
    =
    \frac{\Tr(O)+1}{2}.
    \label{eq:angle-general}
\end{equation}
The complete roto-reflection geometry may be represented by an oriented normal and a signed angle,
\[
    (\bm n,\phi),
\]
with the equivalence
\[
    (\bm n,\phi)\sim(-\bm n,-\phi).
\]

The geometry is covariant under local unitaries. If
\[
    T\mapsto T'=R_A T R_B^T,
\]
then
\begin{equation}
    O\mapsto O'=R_AOR_B^T.
    \label{eq:Ocovariance}
\end{equation}
The particular plane and angle therefore depend on the chosen local Bloch frames. The orientation-reversing character does not:
\[
    \det O=-1.
\]
This is the invariant signature emphasized here.

\section{Partial transposition and negativity}

Let us look at one of the most prominent entanglement criteria, partial transposition (PT) \cite{peres1996, horodecki1996}, and see how it works in the graphical language proposed in this work.

Usually, the story goes as follows. We take the density matrix $\rho$ of a two-qubit state and transpose the matrix entries corresponding to Bob's qubit. If the resulting operator $\rho^\Gamma$ has a negative eigenvalue, the original state is entangled. For two qubits, the converse also holds. A negative eigenvalue means that there is a normalized vector $\ket{\chi}$ in the two-qubit Hilbert space for which
\begin{equation}
    \Tr\!\left(\rho^\Gamma\ket{\chi}\bra{\chi}\right)
    =\bra{\chi}\rho^\Gamma\ket{\chi}<0.
\end{equation}
Such an expectation value is impossible for a physical density matrix, which must be positive semidefinite.

For a generic pure two-qubit state,
\begin{equation}
    \ket{S}=\alpha\ket{00}+\beta\ket{01}
    +\gamma\ket{10}+\delta\ket{11},
\end{equation}
where $|\alpha|^2+|\beta|^2+|\gamma|^2+|\delta|^2=1$, we have $\rho=\ket{S}\bra{S}$, and
\begin{equation}
\rho^\Gamma=
\begin{pmatrix}
|\alpha|^2 & \beta\alpha^* & \alpha\gamma^* & \beta\gamma^*\\
\alpha\beta^* & |\beta|^2 & \alpha\delta^* & \beta\delta^*\\
\gamma\alpha^* & \delta\alpha^* & |\gamma|^2 & \delta\gamma^*\\
\gamma\beta^* & \delta\beta^* & \gamma\delta^* & |\delta|^2
\end{pmatrix}.
\label{eq:generic-pt}
\end{equation}
For general coefficients, neither the negative eigenvalue nor its eigenvector is apparent from this matrix. Finding them directly involves diagonalizing a $4\times4$ Hermitian matrix. Let us see how our graphical representation helps us understand the result and find the negative eigenvector by a much simpler construction.

\begin{figure}[t]
\centering
\begin{tikzpicture}[>=Stealth, scale=0.95]

\def\R{1.0}

\newcommand{\drawtriad}[5]{%
    \begin{scope}[shift={(#1,#2)}]
        \draw[gray!65,thin] (0,0) circle (\R);
        \fill[black] (0,0) circle (1pt);

        \draw[very thick,->,xcol]
            (0,0) -- ({#3*0.74},0)
            node[above] {\scriptsize $x$};

        \draw[very thick,->,ycol]
            (0,0) -- ({#4*(-0.46)},{#4*(-0.40)})
            node[left] {\scriptsize $y$};

        \draw[very thick,->,zcol]
            (0,0) -- (0,{#5*0.78})
            node[right] {\scriptsize $z$};
    \end{scope}
}

\node at (1.8,0) {\large $\bm{\ket{\Phi^+}}$};

\drawtriad{0}{0}{1}{-1}{1}
\drawtriad{3.6}{0}{1}{1}{1}

\node at (0,-1.22) {Alice};
\node at (3.6,-1.22) {Bob};
\node at (1.8,-1.90)
    {$T_{\Phi^+}=\operatorname{diag}(1,-1,1)$};

\node at (1.8,-3.90)
    {\large $\bm{\rho^{\Gamma}}$};

\drawtriad{0}{-4}{1}{-1}{1}
\drawtriad{3.6}{-4}{1}{-1}{1}

\node at (0,-5.22) {Alice};
\node at (3.6,-5.22) {Bob};
\node at (1.8,-5.60) {$T^\Gamma_{\Phi^+}=I_3$};

\node at (1.8,-8.1) {\large $\bm{\ket{\Psi^-}}$};

\drawtriad{0}{-8}{1}{-1}{1}
\drawtriad{3.6}{-8}{-1}{1}{-1}

\node at (0,-9.22) {Alice};
\node at (3.6,-9.22) {Bob};
\node at (1.8,-9.60) {$T_{\Psi^-}=-I_3$};

\draw[thick,->] (5.40,-0.9) -- (5.40,-3.05);
\node[right,align=left] at (5.55,-1.95)
    {\small PT on Bob\\
     \small $y\rightarrow -y$};

\draw[thick,->] (5.40,-4.9) -- (5.40,-7.05);
\node[right,align=left] at (5.55,-5.95)
    {\small Maximally\\
     \small opposing state};

\end{tikzpicture}

\caption{Geometric interpretation of partial
transposition and negativity. Partial transposition
reverses Bob's $y$-axis, producing three aligned
correlations. The singlet opposes all three.}
\label{fig:pt-triptych}
\end{figure}

Let us move analogously to our previous discussions and look at an example of a maximally entangled state in the Schmidt frame, namely $\ket{\Phi^+}=(\ket{00}+\ket{11})/\sqrt{2}$. The Bloch vectors are dots in the middle of the spheres, and the only geometry is given by the relative arrows representing the correlations. The $Z$ and $X$ directions are positively correlated (represented by arrows facing in the same direction on both Bloch spheres), while the $Y$ directions are negatively correlated, $T_{\Phi^+}=\diag(1,-1,1)$. Hence the rotoreflection plane is the \(zx\) plane. Partial transposition of Bob's qubit implies reversing the $Y$ direction on that qubit, and hence destroys the rotoreflection. The resulting correlation structure is now that of the identity, $T^\Gamma_{\Phi^+}=I_3$: every direction on A is maximally correlated with the same direction on B. Let us now see how this implies a negative eigenvalue.

The trace of the product of two operators is a dot product in the space of two-qubit Pauli operators. In our graphical language, we are therefore looking for a state whose correlation arrows point as far as possible opposite to those of $\rho^\Gamma$. A correlation arrow is not a simple arrow like a Bloch vector. It corresponds to an operator $(\mathbf{u}\cdot\boldsymbol{\sigma})\otimes(\mathbf{v}\cdot\boldsymbol{\sigma})$, with unit vectors $\mathbf{u}$ and $\mathbf{v}$, and connects a direction on A with the direction on B maximally correlated with it. It is thus a single, relative object extended across both Bloch spheres. In our visualization, the arrows on A are fixed along $x$, $y$ and $z$, so all the correlation information sits in Bob's arrows, and comparing two states amounts to comparing their arrows on B. For $T^\Gamma_{\Phi^+}=I_3$, the arrows that maximally oppose those of $\rho^\Gamma$ are obtained by inverting all three arrows on Bob's sphere, which gives
\begin{equation}
    T_\chi=\diag(-1,-1,-1)=-I_3.
\end{equation}
We have encountered this geometry before: it is the correlation tensor of the singlet $\ket{\Psi^-}=(\ket{01}-\ket{10})/\sqrt{2}$. In Sec.~3, the singlet appeared as the maximally symmetric ERRP configuration, for which every reflection plane is possible. Here it appears in a different role, as the physical state whose correlations maximally oppose those of the partially transposed $\ket{\Phi^+}$. Its overlap with $\rho^\Gamma$ is indeed negative:

\begin{equation}
    \bra{\Psi^-}\rho^\Gamma\ket{\Psi^-}
    =\frac14(1-3)=-\frac12.
\end{equation}

Having a recipe for finding the eigenvector with a negative eigenvalue, let us see how it works for a partially entangled Schmidt state. The geometry of the entangled layer is the same as before, while the Bloch vectors of the inner sphere face along the $z$ axis; hence they are not modified by PT. But even for a generic separable state, Bob's Bloch vector would be reflected in the \(zx\) plane, which would still give a valid Bloch vector, and hence a valid state overall, without any negativity. In fact, one can also make a visual argument for why a separable state cannot have a negative overlap; we leave this exercise to the reader. Therefore, the negative contribution originates in the entanglement shell, which has the geometry of $\ket{\Phi^+}$. Hence the state that makes the partial transpose most negative is again $\ket{\Psi^-}$.

Moving on to arbitrary pure two-qubit states, let us formulate the rule connecting the state that is partially transposed with the eigenvector of $\rho^\Gamma$ with the negative eigenvalue. Given the state, take only the outer shell representing entanglement (mathematically, this amounts to extracting orthogonal factor $O$ from the correlation tensor). Then rotate all of Bob's arrows by $\pi$ around the axis of transposition. The resulting arrows represent the negative eigenvector. This follows from the observation that the reflection of a single axis (PT), followed by the inversion of all three axes (maximal negative overlap), is just a rotation by $\pi$ around that axis. 

\begin{equation}
    T_\chi
    = O\,O_0\,(-I_3)
    = O\,R_y(\pi).
\end{equation}

The eigenvector that remained hidden in the generic matrix of Eq.~\eqref{eq:generic-pt} is thus obtained by a simple graphical construction, without diagonalizing $\rho^\Gamma$.

Having considered this short example of the intuition behind negativity, let us try to reinterpret the rotoreflection geometry of entanglement. The orientation-reversing character of the ERRP can be understood as a geometric manifestation of quantum-state positivity. Partial transposition removes this reversal, exposing a pattern of correlations that cannot belong to a physical state. 

In the arrow language, correlations perpendicular to the Bloch-vector axis cannot be increased independently: additional correlation along one transverse axis must be accompanied by an equal anticorrelation along the other. This is clearest in the Schmidt frame, where $z$ is the Bloch-vector axis and the correlations along the transverse directions enter the correlation tensor as 
\begin{equation} 
T=\diag(C,-C,1). 
\end{equation} 
The two transverse contributions therefore cancel in the signed correlation balance, leaving $1+C-C=1$. Partial transposition reverses the $y$ direction on Bob's side and changes the tensor to 
\begin{equation} 
T^\Gamma=\diag(C,C,1). 
\end{equation}
What was compensation now becomes reinforcement. Evaluating the partially transposed state on the singlet gives 
\begin{equation} 
\bra{\Psi^-}\rho^\Gamma\ket{\Psi^-} =\frac14(1-C-C-1) =-\frac{C}{2}. 
\end{equation} 
Thus, the negative eigenvalue appears precisely when PT turns the pair $C,-C$, whose opposite signs preserve positivity, into the pair $C,C$, whose contributions add with the same negative sign in the overlap. Negativity is therefore the algebraic trace of a broken geometrical balance.

\section{Conclusion}

In this work, we have provided an intuitive geometric picture of pure two-qubit states, built on their Pauli decomposition. At its centre stands a single geometric object, the ERRP, which organizes the directional structure of the correlations. The procedure that extracts this geometry from the algebra is explicit, so researchers may use it to create visualizations of their own.

The value of this construction lies not only in visualization, but also in geometric reasoning. It allows us to follow how single- and double-qubit unitaries transform the correlations, see the degree of entanglement through their contraction, and develop an intuition for the overlap between different states. The connection to partial transposition provides a particularly striking example. The orientation reversal underlying the ERRP acquires a physical meaning as a manifestation of the positivity constraint on pure-state correlations. Partial transposition removes this reversal, and the state revealing the resulting negative eigenvalue can be constructed directly from the original geometry by a simple rotation.

More generally, the geometric picture brings together several features that are usually discussed separately. The ERRP, the correlation structure of Schmidt states, and the emergence of negativity under partial transposition all point toward the same theme: physical quantum correlations appear to obey a balance between correlation and anticorrelation, which partial transposition breaks for every entangled pure state. The same $1/3$ boundary appears both in isotropic positive correlations, where three agreeing correlations of strength $1/3$ exactly exhaust the balance, and in Werner-state separability. This raises the possibility that the balance extends beyond pure states. Whether this reflects a deeper geometric constraint or merely a useful way of organizing known facts remains an open question. Another natural direction is to develop the graphical intuition for overlaps into a more general geometric calculus.

This work provides a new way of organizing the geometry of bipartite quantum states. The Bloch sphere owes much of its success to the fact that it turns an abstract quantum state into a directly visible geometric object. The present construction attempts to do something similar for two qubits: to separate local properties from relational ones, and to represent the latter in a form that can be inspected and reasoned about. For a single qubit, one arrow tells the whole story. For two qubits, the picture becomes two Bloch arrows and a mirror.

\section*{Acknowledgements}
S.F.\ acknowledges support from Project No.\ 1.1.1.8/1/24/I/003, ``Strengthening the research and development capacity of doctoral studies at the University of Latvia in smart specialization areas.'' S.F.\ is also grateful to the Yukawa Institute for Theoretical Physics, Kyoto University, where part of this work was carried out during the workshop YITP-T-25-01.

\bibliographystyle{unsrt}
\bibliography{references}

\end{document}